\documentclass[conference]{IEEEtran}
\IEEEoverridecommandlockouts
\usepackage{cite}
\usepackage{amsmath,amssymb,amsfonts}
\usepackage{algorithmic}
\usepackage{graphicx}
\usepackage{textcomp}
\usepackage{xcolor}

\def\BibTeX{{\rm B\kern-.05em{\sc i\kern-.025em b}\kern-.08em
    T\kern-.1667em\lower.7ex\hbox{E}\kern-.125emX}}
\begin{document}
\title{Energy Efficiency of Generalized Spatial Modulation Aided Massive MIMO Systems\\
\thanks{This work was supported in part by the National Key Research and Development Program of China under Grant 2017YFE0121600.}
}

\author{\IEEEauthorblockN{Shuang Zheng\IEEEauthorrefmark{1}, Jing Yang\IEEEauthorrefmark{1}, Xiaohu Ge\IEEEauthorrefmark{1}, Yonghui Li\IEEEauthorrefmark{2}, Lin Tian\IEEEauthorrefmark{3},
Jinglin Shi\IEEEauthorrefmark{3}\IEEEauthorrefmark{4}}
\IEEEauthorblockA{\IEEEauthorrefmark{1}\textit{School of Electronic Information and Communications}\\
\textit{Huazhong University of Science and Technology}, Wuhan, Hubei, China \\}
\IEEEauthorblockA{\IEEEauthorrefmark{2}\textit{School of Electrical and Information Engineering}\\
\textit{ University of Sydney}, Sydney, Australia\\}
\IEEEauthorblockA{\IEEEauthorrefmark{3}\textit{Beijing Key Laboratory of Mobile Computing and Pervasive Devices}\\
\textit{Institute of Computing Technology, Chinese Academy of Sciences}, China\\}
\IEEEauthorblockA{\IEEEauthorrefmark{4}\textit{University of Chinese Academy of Sciences}, China\\}
Contact Email: xhge@mail.hust.edu.cn
}

\maketitle

\begin{abstract}
One of focuses in green communication studies is the energy efficiency (EE) of massive multiple-input multiple-output (MIMO) systems. Although the massive MIMO technology can improve the spectral efficiency (SE) of cellular networks by configuring a large number of antennas at base stations (BSs), the energy consumption of radio frequency (RF) chains increases dramatically. The increment of energy consumption is caused by the increase of RF chain number to match the antenna number in massive MIMO communication systems. To overcome this problem, a generalized spatial modulation (GSM) solution is presented to simultaneously reduce the number of RF chains and maintain the SE of massive MIMO communication systems. A EE model is proposed to estimate the transmission and computation power of massive MIMO communication systems with GSM. Simulation results demonstrate that the EE of massive MIMO communication systems with GSM outperforms the massive MIMO communication systems without GSM. Besides, the computation power consumed by massive MIMO communication systems with GSM is effectively reduced.
\end{abstract}


\section{Introduction}
Having hundreds of antennas at base stations (BSs), the massive multiple-input multiple-output (MIMO) technology can improve the spatial diversity and array gain to increase the spectral efficiency (SE) \cite{1Lu},\cite{2Ge}. The conventional MIMO technology configures a separate radio frequency (RF) chain for each antenna. For massive MIMO communication systems with hundreds of antennas, the solution with one RF chain corresponding to one antenna will greatly increase the energy consumed by RF chains at BSs \cite{3Bjornson}. To overcome the great power consumption due to the increase of RF chain number, finding new solutions for massive MIMO communications systems with limited RF chain number is an emerging challenge for the fifth generation (5G) mobile communication systems.

For conventional mobile communication systems, the power consumption is divided into the transmission power, computation power and the other power consumption\cite{4Xiang}. The transmission power mainly represents the energy consumed by power amplifier (PA) and RF chains; the computation power is mainly consumed by baseband units (BBUs); the other power consumption mainly includes the power consumption of active cooling at BSs \cite{5Auer}. In conventional mobile communication systems, the power consumption of PA is the largest part in the total power consumed at BSs, and the computation power is usually ignored or set to a fixed value \cite{6Desset}. However, in 5G mobile communication systems, the transmission rate can be greatly improved by both the massive MIMO and millimeter wave (mmWave) technologies at BSs \cite{7Andrews}. Moreover, due to the deployment of ultra-dense small cell networks, the distance between customers and BSs is obviously reduced\cite{8Ge},\cite{9X}. The reduced distances lead to a decrease of transmission power in 5G mobile communication systems\cite{10XGe}. Meanwhile, with the rapid growth of traffic, more computation power is consumed by the signal processing in BBUs of 5G mobile communication systems. The computation power is accounted for more than 50\% of the total power consumption at 5G BSs \cite{11Ge}. In this case, the computation power plays a dominant role in the total power consumption of 5G mobile communication systems. Thus, it is crucial to come up with strategies of reducing the computation power in massive MIMO communication systems.

Generalized spatial modulation (GSM) technology is emerging as a potential solution to maintain the SE of massive MIMO communication systems and reduce the RF chain number simultaneously\cite{12Wang}. Different from the conventional MIMO technology, the GSM technology transmits the symbols by the activated antennas, which is selected by the space-domain information. Thus the information is simultaneously transmitted by the amplitude phase modulation (APM) symbols and the indices of activated antennas \cite{12Wang}. Combing GSM and massive MIMO technologies, the amount of RF chains is decreased and then the power consumption of RF chains is greatly reduced \cite{13Renzo},\cite{14X},\cite{15XGe}. Therefore, a higher energy efficiency (EE) can be achieved by combining the massive MIMO technology with GSM. A simple massive MIMO communication system with GSM was investigated for indoor transmissions over line-of-sight (LoS) channels and the SE was analyzed in \cite{16He}. A detection algorithm was proposed to improve the EE of massive MIMO communication systems with GSM in multi-cell multi-user scenarios \cite{17Patchar}. The proportion of computation power in the total power consumption obviously increased in 5G mobile communication systems \cite{11Ge}. However, most studies related to the EE of massive MIMO communication systems with GSM have not considered the computation power, which leads to the contributions of this paper.

In this paper, the EE of massive MIMO communication systems with GSM is analyzed for a single-cell multi-user scenario. Moreover, the computation power consumption model of massive MIMO communication systems with GSM is proposed. Simulation results show that the EE of massive MIMO communication systems with GSM outperforms the massive MIMO communication systems without GSM. Besides, the power of computation consumed by massive MIMO communication systems with GSM is effectively reduced.

The content of this paper is constructed as follows. Section II describes a massive MIMO communication system with GSM. The capacity and total power consumption model of massive MIMO communication systems with GSM are derived in Section III. Simulations results and analysis are presented in Section IV, and the Section V concludes this paper.

\section{System Model}
\begin{figure}[htbp]
\centerline{\includegraphics[width=0.5\textwidth]{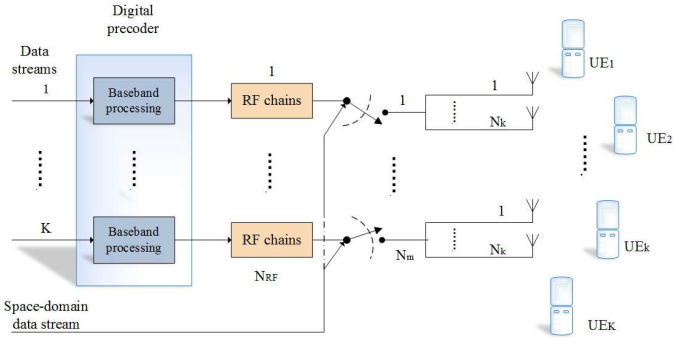}}
\caption{The massive MIMO communication system with GSM.}
\label{fig1}
\end{figure}

In Fig.~\ref{fig1}, we depict a massive MIMO communication system with GSM, where $N_\text{T}$ transmit antennas and $K$ single-antenna users are configured. As described in Fig.~\ref{fig1}, the input data streams are divided into $K$ data streams and an extra space-domain data stream. The digital precoder, which consists of $K$ baseband processing units, plays the role of allocating power. After the digital baseband processing, $N_\text{RF}$ RF symbols are produced.

The $N_\text{T}$  transmit antennas are grouped into $N_\text{m}$ antenna groups and each antenna group has $N_\text{k}$ transmit antennas, i.e., ${N}_\text{T}={N}_\text{m}\times {N}_\text{k}$. Considering digital precoder design constraints, ${N}_\text{m}\ge {N}_\text{RF}$ should be satisfied in massive MIMO communication systems with GSM. After data streams processed by RF chains, the spatial-domain data stream randomly select ${N}_\text{RF}$ from ${N}_\text{m}$ antenna groups to transmit symbols based on the GSM technology. In this case, the remaining $({N}_\text{m}-{N}_\text{RF} )$ antenna groups are inactive.

The space-domain data stream plays the role of specifying the combination of activated antenna groups. Thus the total number of valid combination of antenna groups is given by \cite{12Wang}

\begin{equation}
M={{2}^{\left\lfloor {{\log }_{2}}\left( _{{{{N}}_\text{RF}}}^{{{{N}}_\text{m}}} \right) \right\rfloor }}.
\label{eq1}
\tag{1}
\end{equation}
The ${m}$-th activated antenna group combination is denoted by ${{\mathbf{u}}_{m}}\triangleq {{\left[ {{{u}}_{m1}},{{{u}}_{m2}},\cdots ,{{{u}}_{m{{{N}}_\text{RF}}}} \right]}^{{T}}}$, $( m=1, 2, \cdots , M)$, which follows the restriction $1\le {{{u}}_{m1}}\le {{{u}}_{m2}}\le \cdots \le {{{u}}_{m{{{N}}_\text{RF}}}}\le {{{N}}_\text{m}}$. Therefore, the ${m}$-th activated antenna group matrix ${{\mathbf{C}}_m}\in {{\mathbb{C}}^{{{{N}}_\text{T}}\times {{{N}}_\text{RF}}}}$ denoting the GSM matrix is expressed as \cite{18He}

\begin{equation}
{{\mathbf{C}}_{m}}\triangleq [{{\mathbf{e}}_{{{{u}}_{m1}}}},{{\mathbf{e}}_{{{{u}}_{m2}}}},\cdots {\mathbf{e}}_{{{{u}}_{mN_\text{RF}}}}]\otimes {{\mathbf{1}}_{{{{N}}_\text{k}}}}.
\label{eq2}
\tag{2}
\end{equation}
Note that the transmit antennas are grouped, the group transmission power is expanded ${{{N}}_\text{k}}$ times compared with the single antenna transmission power. In order to normalize the power of group transmit antennas, the GSM matrix is adjusted by

\begin{equation}
{{\mathbf{C}}_m}\triangleq \sqrt{\frac{1}{{{{N}}_\text{k}}}}[ {{\mathbf{e}}_{{{{u}}_{m1}}}},{{\mathbf{e}}_{{{{u}}_{m2}}}},\cdots ,{{\mathbf{e}}_{{{{u}}_{mN_\text{RF}}}}}]\otimes {{\mathbf{1}}_{{{{N}}_\text{k}}}}.
\label{eq3}
\tag{3}
\end{equation}

After processed by the digital precoding and GSM technology, the symbols are transmitted to users through the channel matrix of activated antenna groups. The channel matrix is given by ${{\mathbf{H}}^{{H}}}={{[ {{\mathbf{h}}_{1}},\ldots ,{{\mathbf{h}}_k},\ldots ,{{\mathbf{h}}_K} ]}^{{H}}}\in {{\mathbb{C}}^{{K}\times {{{N}}_\text{T}}}}$, where ${{\mathbf{h}}_k}\in {{\mathbb{C}}^{{{{N}}_\text{T}}\times 1}}$ is the instantaneous propagation channel among the $k$-th user and transmit antennas. To simplify the derivation in this paper, a Rayleigh channel model is adopted for massive MIMO communication systems \cite{19Gao}

\begin{equation}
{{\mathbf{h}}_k}\sim \mathcal{C}\mathcal{N}( {{\mathbf{0}}_{{{{N}}_\text{T}}}},l({{d}_k} ){{\mathbf{I}}_{{{{N}}_\text{T}}}}),
\label{eq4}
\tag{4}
\end{equation}
where $l(\mathbf{d})\!=\!\frac{\overline{d}}{{{\left\| \mathbf{d} \right\|}^{\alpha}}}$ is a small-scale fading distribution dominated by the channel attenuation,  $\mathbf{d}\!=\!{{[{{d}_{1}},\ldots ,{{d}_k},\ldots ,{{d}_K}]}^{T}}\in {{\mathbb{C}}^{K\times 1}}$ represents the distances among ${K}$ users and transmit antennas. ${d}_{k}$ follows the constraint ${{d}_{\text{min}}}\le {{d}_{k}}\le {{d}_{\text{max}}}$, where ${{d}_{\text{max}}}$ is the maximum distance between users and transmit antennas, ${{d}_{\text{min}}}$ is the minimum distance among users and transmit antennas. $\overline{d}$ is the channel attenuation at ${d}_{\min}$, $\alpha$ is the path loss coefficient \cite{3Bjornson}. The digital precoding matrix is given by
$\mathbf{B}=\left[ {{\mathbf{b}}_{1}},\cdots {{\mathbf{b}}_k},\cdots ,{{\mathbf{b}}_K} \right]\in {{\mathbb{C}}^{{{{N}}_\text{RF}}\times K}}$, where ${{\mathbf{b}}_k}\in {{\mathbb{C}}^{{{{N}}_\text{RF}}\times 1}}$ is the precoding vector of the $k$-th user. In this paper, a zero-forcing precoder is configured for the digital precoder, which is expressed as

\begin{equation}
\mathbf{B}_\text{ZF}={{\beta }_\text{ZF}}{{\left( \mathbf{H}{{\mathbf{C}}_m} \right)}^{{H}}}{{\left( \left( \mathbf{H}{{\mathbf{C}}_m} \right){{\left( \mathbf{H}{{\mathbf{C}}_m} \right)}^{{H}}} \right)}^{-1}},
\label{eq5}
\tag{5}
\end{equation}
where ${{\beta }_\text{ZF}}=\sqrt{\frac{{{{P}}_{\max }}}{tr{\left( {\left( \left( \mathbf{H}{{\mathbf{C}}_m} \right){\left( \mathbf{H}{{\mathbf{C}}_m} \right)}^{H} \right)}^{-1} \right)}}}$ is a normalization factor.

The signal received by the $k$-th  user is formulated as
\begin{equation}
{{{y}}_k}=\mathbf{h}_k^{{H}}{{\mathbf{C}}_m}{{\mathbf{b}}_k}\mathbf{x}+{{{n}}_k},
\label{eq6}
\tag{6}
\end{equation}
where $\mathbf{x}\sim \mathcal{C}\mathcal{N}\left( \mathbf{0},{{\mathbf{I}}_K} \right)$ is the Gaussian-distributed input signal vector, $\mathbf{n}\sim \mathcal{C}\mathcal{N}\left( \mathbf{0},\sigma _\text{N}^{2}{{\mathbf{I}}_K} \right)$ is the additive white Gaussian noise.

\section{Energy Efficiency of massive MIMO communication systems with GSM}

\subsection{Energy Efficiency Model}

Without loss of generality, in this paper the EE model is expressed as the ratio of capacity to total power consumption in massive MIMO communication systems with GSM,

\begin{equation}
\text{ }\!\!\eta\!\!\text{ }=\frac{{{{R}}_\text{total}}}{{{{P}}_\text{total}}},
\label{eq7}
\tag{7}
\end{equation}
where ${{{R}}_\text{total}}$ is the capacity of massive MIMO communication systems with GSM, ${{{{P}}_\text{total}}}$ is the total power consumption of massive MIMO communication systems with GSM.

\subsection{Capacity of Massive MIMO Communication Systems with GSM}

Generally, the SE of mobile communication systems is expressed as the ratio of capacity to bandwidth. In this paper, the mutual information is used to quantify the SE between the received signal ${y}_k$ of the $k$-th user and input signal $\mathbf{x}$ \cite{18He},

\begin{equation}
R={I}\left( {{y}_k};\mathbf{x},{m} \right)=I\left( {{y}_k};\mathbf{x}|m \right)+I({{y}_k};{m}),
\label{eq8}
\tag{8}
\end{equation}
where ${I}\left( {{y}_{k}};\mathbf{x}|{m} \right)$ is the APM-domain mutual information after antenna groups combination is selected, $I\left( {{y}_{k}};{m} \right)$ is the mutual information between the received signal ${{y}_{k}}$ of the $k$-th user and antenna groups combination index ${m}$. Based on the definition of mutual information \cite{20An}, we can obtain

\begin{equation}
I\left( {{y}_{k}};\mathbf{x}|m \right)=H\left( {{y}_{k}}|m \right)-H\left( {{y}_{k}}|\mathbf{x},m \right),
\label{eq9a}
\tag{9a}
\end{equation}
with
\begin{equation}
H\left( {{y}_{k}}|m \right)=\frac{1}{M}\sum\limits_{m=1}^{M}{\left[ {{\log }_{2}}\left( \pi e \right)+{{\log }_{2}}\left( \left| {{\Sigma }_{m}} \right| \right) \right]},
\label{eq9b}
\tag{9b}
\end{equation}

\begin{equation}
\begin{gathered}
H\left( {{y}_{k}}|\mathbf{x},m \right)=-{{E}_{{{y}_{k}},\mathbf{x},m}}\left\{ {{\log }_{2}}\mathcal{P}\left( {{y}_{k}}|\mathbf{x},m \right) \right\}
\hfill \\
\;\;\;\;\;\;\;\;\;\;\;\;\;\;\;\;\;\;\;=\frac{1}{M}\sum\limits_{m=1}^{M}{\left[ {{\log }_{2}}\left( \pi e \right)+{{\log }_{2}}\left( \left| \sigma _{\text{N}}^{2} \right| \right) \right]}. \hfill \\
\label{eq9c}
\tag{9c}
\end{gathered}
\end{equation}
Moreover, the APM-domain mutual information after selecting antenna groups combination is simplified as

\begin{equation}
I\left( {{y}_{k}};\mathbf{x}|m \right)=\frac{1}{M}\sum\limits_{m=1}^{M}{{{\log }_{2}}\left( \left| \frac{{{\Sigma }_{m}}}{\sigma _{\text{N}}^{2}} \right| \right)}.
\label{eq10}
\tag{10}
\end{equation}

Similarly, $I\left( {{y}_{k}};m \right)$ is expressed as

\begin{equation}
I\left( {{y}_{k}};m \right)=\frac{1}{M}\sum\limits_{m=1}^{M}{\int{\mathcal{P}\left( {{y}_{k}}|n \right){{\log }_{2}}\left[ \frac{\mathcal{P}\left( {{y}_{k}}|n \right)}{\frac{1}{M}\sum\limits_{t=1}^{M}{\mathcal{P}\left( {{y}_{k}}|t \right)}} \right]}}d{{y}_{k}},
\label{eq11}
\tag{11}
\end{equation}
which is simplified as

\begin{equation}
I\left( {{y}_{k}};m \right)\approx {{\log }_{2}}\frac{M}{2}-\frac{1}{M}\sum\limits_{n=1}^{M}{{{\log }_{2}}\left( \sum\limits_{t=1}^{M}{\frac{\left| {{\Sigma }_{n}} \right|}{\left| {{\Sigma }_{n}}+{{\Sigma }_{t}} \right|}} \right)}.
\label{eq12}
\tag{12}
\end{equation}

Therefore, the SE of $k$-th user in massive MIMO communication systems with GSM is derived as

\begin{equation}
\begin{gathered}
{{{R}}_k}=\frac{1}{{M}}\sum\limits_{{m}=1}^{{M}}{{{\log }_{2}}\left( \left| \frac{1}{\sigma _\text{N}^{2}}{{\sum }_m} \right| \right)}+{{\log }_{2}}\left( \frac{{M}}{2} \right) \hfill \\
\;\;\;\;\;-\frac{1}{{M}}\sum\limits_{n=1}^{{M}}{{{\log }_{2}}\left( \sum\limits_{{t}=1}^{{M}}{\frac{\left| {{\sum }_n} \right|}{\left| {{\sum }_n}+{{\sum }_t} \right|}} \right)}, \hfill \\
\label{eq13}
\tag{13}
\end{gathered}
\end{equation}
where ${{\Sigma }_m}$ is the covariance matrix of ${{y}_{k}}$ considering the $m$-th activated antenna groups \cite{18He}

\begin{equation}
{{\sum}_m}=\sigma_\text{N}^{2}+\mathbf{h}_k^{{H}}{{\mathbf{C}}_m}{{\mathbf{b}}_k}
{\mathbf{b}}_k^{H}\mathbf{C}_m^{{H}}{{\mathbf{h}}_k}.
\label{eq14}
\tag{14}
\end{equation}

As a consequence, the capacity of massive MIMO communication systems with GSM is derived by

\begin{equation}
{{{R}}_\text{total}}={W}\sum\limits_{{k}=1}^{{K}}{{{{R}}_k}},
\label{eq15}
\tag{15}
\end{equation}
where ${W}$ is the bandwidth.

To compare with massive MIMO communication systems with GSM, the capacity of conventional massive MIMO communication systems is expressed as

\begin{equation}
{{{R}}_k}={W}{{\log }_{2}}\left( 1+\frac{{{\mathbf{h}}_k}{{\mathbf{b}}_k}\mathbf{b}_k^{{H}}\mathbf{h}_k^{{H}}}{\sum\limits_{{i}=1,{i}\ne {k}}^{{K}}{{{\mathbf{h}}_k}{{\mathbf{b}}_i}\mathbf{b}_i^{H}\mathbf{h}_k^{H}+\sigma _\text{N}^{2}}} \right).
\label{eq16}
\tag{16}
\end{equation}

\subsection{Total Power Consumption Model}
 In this paper the power consumption of massive MIMO communication systems with GSM is classified into the transmission power, computation power and the fixed power.

\subsubsection{Transmission Power}

The transmission power ${{{P}}_\text{T}}$ includes three parts: the energy consumed by PA ${{{P}}_\text{PA}}$, the energy consumed by RF chains ${{{P}}_\text{RF\_chains}}$, and the energy consumed by switches ${{{P}}_\text{switch}}$. The transmission power is given by

\begin{equation}
{{{P}}_\text{T}}={{{P}}_\text{PA}}+{{{P}}_\text{RF\_chains}}+{{{P}}_\text{switch}}.
\label{eq17}
\tag{17}
\end{equation}
The energy consumed by PA ${{{P}}_\text{PA}}$ is calculated by

\begin{equation}
{{{P}}_\text{PA}}=\frac{{{{P}}_{\max }}}{\text{ }\!\!\gamma\!\!\text{ }},
\label{eq18}
\tag{18}
\end{equation}
where $\gamma$ is the exchange of PA. Note that the power allocation vector is simultaneously depended on the space-domain data stream and instantaneous channel state information (CSI). Besides, the maximum power of receiver is less than or equal to a fixed value $P_\text{max}$. The energy consumed by RF chains is calculated by\cite{21Auer}

\begin{equation}
{{{P}}_\text{RF\_chains}}={{{N}}_\text{RF}}\times {{{P}}_\text{RF}},
\label{eq19}
\tag{19}
\end{equation}
where ${{{P}}_\text{RF}}$ is the energy consumed by each RF chain. ${{{P}}_\text{switch}}$ is the energy consumption of switches when the space-domain data stream selects the activated antenna groups\cite{22Dai}

\begin{equation}
{{{P}}_\text{switch}}={{{N}}_\text{RF}}\times {{{P}}_\text{each\_switch}},
\label{eq20}
\tag{20}
\end{equation}
where ${{{P}}_\text{each\_switch}}$ is the energy consumption of each switch.

In the end, the transmission power is quantified as

\begin{equation}
{{P}_{T}}=\frac{{{{P}}_{\max }}}{\text{ }\!\!\gamma\!\!\text{ }}+{{{N}}_\text{RF}}\times {{{P}}_\text{RF}}+{{{N}}_\text{RF}}\times {{{P}}_\text{each\_switch}}.
\label{eq21}
\tag{21}
\end{equation}

\subsubsection{Computation Power}

In this power consumption model, the computation power is taken into account and cannot be treated as a fixed value. The computation power ${{{P}}_\text{C}}$ includes three parts: the energy consumption for channel estimation ${{{P}}_\text{CE}}$, the energy consumption for channel coding ${{{P}}_\text{CD}}$, and the energy consumption for linear processing ${{{P}}_\text{LP}}$ \cite{23Ge}

\begin{equation}
{{{P}}_\text{C}}={{{P}}_\text{CE}}+{{{P}}_\text{CD}}+{{{P}}_\text{LP}}.
\label{eq22}
\tag{22}
\end{equation}

Without loss of generality, coherent blocks per unit time is assumed by $\frac{{W}}{{U}}$, where ${W}$ represents the bandwidth and ${U}$ represents the coherent block. The CSI estimation of massive MIMO communication systems with GSM based on the pilot is performed once for each block. The proposed system receives the pilot signal which is denoted as a ${{{N}}_\text{T}}\times \tau{K}$ matrix. The user's channel is evaluated by multiplying with the pilot sequence of length $\tau{K}$ \cite{24Mohammed}. Therefore, the energy consumed by channel estimation is calculated by

\begin{equation}
{{{P}}_\text{CE}}=\frac{{W}}{{U}}\frac{2\text{ }\!\!\tau\!\!\text{ }{{{N}}_\text{T}}{{{K}}^{2}}}{{{{L}}_\text{BS}}},
\label{eq23}
\tag{23}
\end{equation}
where ${{{L}}_\text{BS}}$ is the computation efficiency in massive MIMO communication systems with GSM. The power required for channel coding is given by\cite{25Bjornson}

\begin{equation}
{{{P}}_\text{CD}}={{{P}}_\text{COD}}{{{R}}_\text{total}},
\label{eq24}
\tag{24}
\end{equation}
where ${{{P}}_\text{COD}}$ is the coding power (in Watt per bit/s).

The power consumed by linear processing is expressed as

\begin{equation}
\begin{gathered}
{{{P}}_\text{LP}}=\frac{{W}}{{U}}\frac{1}{{{{L}}_\text{BS}}}\left( 16{{{K}}^{2}}{{{N}}_\text{RF}}+12{{{K}}^{3}}+8{{{N}}_\text{RF}}{K} \right)\hfill \\
\;\;\;\;\;\;\;+\frac{{W}}{{{{L}}_\text{BS}}}\times 8{{{N}}_\text{RF}}{K},\hfill \\
\end{gathered}
\label{eq25}
\tag{25}
\end{equation}
where the first item $\frac{{W}}{{U}}\frac{1}{{{{L}}_\text{BS}}}\left( 16{{{K}}^{2}}{{{N}}_\text{RF}}+12{{{K}}^{3}}+8{{{N}}_\text{RF}}{K} \right)$ is the power required for the computation of zero-forcing precoding matrix. The precoding matrix is calculated once in each coherent block. The zero-forcing precoding requires $16{{{K}}^{2}}{{{N}}_\text{RF}}+12{{{K}}^{3}}+8{{{N}}_\text{RF}}{K}$ floating-point operations. The second item $\frac{{W}}{{{{L}}_\text{BS}}}\times 8{{{N}}_\text{RF}}{K}$ is the energy consumption of conducting one matrix-vector multiplication for each symbol.

Therefore, the computation power is quantified as

\begin{equation}
\begin{gathered}
{{{P}}_\text{C}}=\frac{{W}}{{U}}\frac{2\text{ }\!\!\tau\!\!\text{ }{{{N}}_\text{T}}{{{K}}^{2}}}{{{{L}}_\text{BS}}}+{{{P}}_\text{COD}}{{{R}}_\text{total}} \hfill \\
\;\;\;\;\;+\frac{{W}}{{U}}\frac{1}{{{{L}}_\text{BS}}}\left( 16{{{K}}^{2}}{{{N}}_\text{RF}}+12{{{K}}^{3}}+8{{{N}}_\text{RF}}{K} \right) \hfill \\
\;\;\;\;\;+\frac{{W}}{{{{L}}_\text{BS}}}\times 8{{{N}}_\text{RF}}{K}. \hfill \\
\end{gathered}
\label{eq26}
\tag{26}
\end{equation}

\subsubsection{Fixed Power}

The fixed power ${{{P}}_\text{FIX}}$ is a constant quantity mainly including the energy consumed by site-cooling and control signaling, which is configured as ${{{P}}_\text{FIX}}=1{\text{W}}$ \cite{5Auer}.

In the end, the total power consumption of massive MIMO communication systems with GSM is quantified as

\begin{equation}
\begin{gathered}
{{{P}}_\text{total}}={{{P}}_\text{T}}+{{{P}}_\text{C}}+{{{P}}_\text{FIX}}
\hfill \\
\;\;\;\;\;\;\;\;=\frac{{{{P}}_{\max }}}{\text{ }\!\!\gamma\!\!\text{ }}+{{{N}}_\text{RF}}\times {{{P}}_\text{RF}}+{{{N}}_\text{RF}}\times {{{P}}_\text{each\_switch}} \hfill \\
\;\;\;\;\;\;\;\;+\frac{{W}}{{U}}\frac{2\text{ }\!\!\tau\!\!\text{ }{{{N}}_\text{T}}{{{K}}^{2}}}{{{{L}}_\text{BS}}}+{{{P}}_\text{COD}}{{{R}}_\text{total}}+\frac{{W}}{{{{L}}_\text{BS}}}\times 8{{{N}}_\text{RF}}{K}+1 \hfill \\
\;\;\;\;\;\;\;\;+\frac{{W}}{{U}}\frac{1}{{{{L}}_\text{BS}}}\left( 16{{{K}}^{2}}{{{N}}_\text{RF}}+12{{{K}}^{3}}+8{{{N}}_\text{RF}}{K} \right). \hfill \\
\end{gathered}
\label{eq27}
\tag{27}
\end{equation}

\section{Simulation Results and Discussions}
In this section, simulations are performed to verify the remarkable performance of massive MIMO communication systems with GSM. Moreover, the computation power is specifically analyzed. Simulation results indicate that the computation power plays a significant role in massive MIMO communication systems with GSM. The default simulation parameters are specified in Table I.

\begin{table}[!htbp]
        \caption{SIMULATION PARAMETERS}\label{tab1}
        \centering
        \begin{tabular}{|c|c|c|}
        \hline
        \textbf{Notations} & \textbf{Descriptions} & \textbf{Values} \\
        \hline
        $N_\text{T}$ & Number of transmit antennas & 128 \\
        \hline
        $N_\text{RF}$ & Number of RF chains & 63 \\
        \hline
        $N_\text{m}$ & Number of antenna groups & 64 \\
        \hline
        $\gamma$ & PA efficiency at the BSs & 0.39 \\
        \hline
        $P_{\text{RF}}$ & Power consumed by the RF chains & 48 mW \\
        \hline
        $W$ & Transmission Bandwidth & 20 MHZ \\
        \hline
        $U$ & Coherence Block & 1800 \\
        \hline
        $\tau$ & Relative Pilot Length & 1 \\
        \hline
        $P_\text{COD}$ & Power required for coding of datas & $1\!\times\!{{10}^{-10}}$ Watt \\
        \hline
        $L_\text{BS}$ & Calculation efficiency at the BS & 12.8 Gflops/W \\
        \hline
        ${{{P}}_\text{each\_switch}}$ & Energy consumed by the single switch & 5 mW \\
        \hline
        $\overline{d}$ & the channel attenuation & ${{10}^{-3.53}}$ \\
        \hline
        $\alpha $ & the pass-loss coefficient & 3.76 \\
        \hline
        \end{tabular}
        \label{table1.}
 \end{table}

Fig. 2 compares the total power consumption of massive MIMO communication systems with GSM and without GSM under different numbers of users. In Fig. 2, the total power consumption increases as the number of users increases. The massive MIMO communication systems with GSM consumes less power than conventional massive MIMO communication systems as the number of users is constant.

Fig. 3 depicts the computation power of BSs under different numbers of users. It is seen that the computation power increases with the increase of the number of users. The reason is that the computation power for channel estimation and linear processing is in proportion to the number of users. Moreover, the computation power of massive MIMO communication systems with GSM is always less than that of conventional massive MIMO systems as the number of users is constant. Based on the consequences in Fig. 2 and Fig. 3, it is clearly observed that the total power consumption and computation power are reduced by massive MIMO communication systems with GSM. Besides, the computation power is accounted for more than 50\% in the total power consumption, which can not be ignored or treated as a fixed value in 5G mobile communication systems.

 \begin{figure}[htbp]
\centerline{\includegraphics[width=0.5\textwidth]{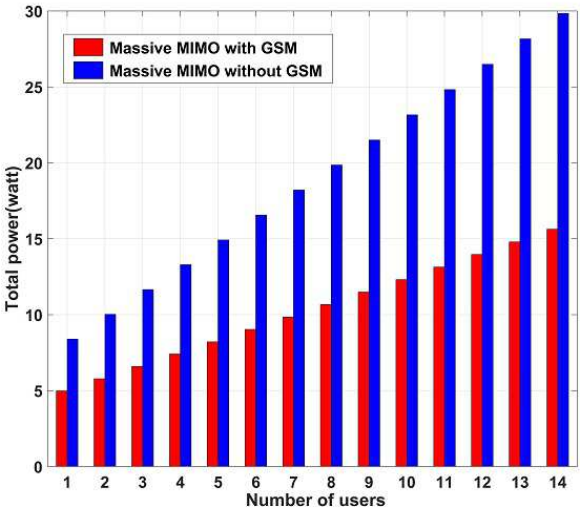}}
\caption{The total power of BSs under different numbers of users.}
\label{fig2}
\end{figure}

\begin{figure}[htbp]
\centerline{\includegraphics[width=0.5\textwidth]{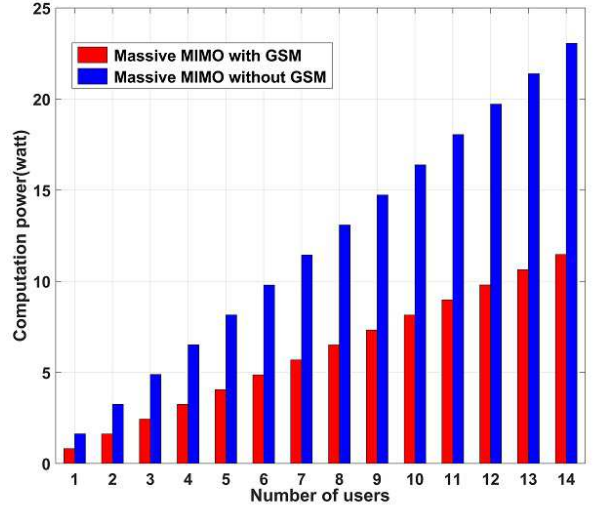}}
\caption{The computation power of BSs under different numbers of users.}
\label{fig3}
\end{figure}

\begin{figure}[htbp]
\centerline{\includegraphics[width=0.5\textwidth]{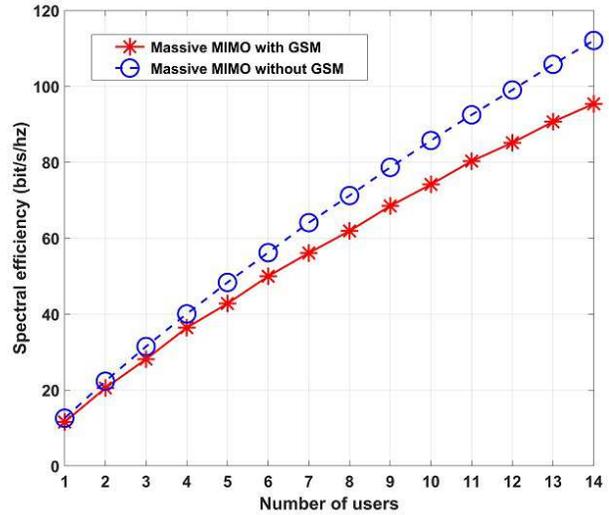}}
\caption{Spectral efficiency under different numbers of users.}
\label{fig4}
\end{figure}

\begin{figure}[htbp]
\centerline{\includegraphics[width=0.5\textwidth]{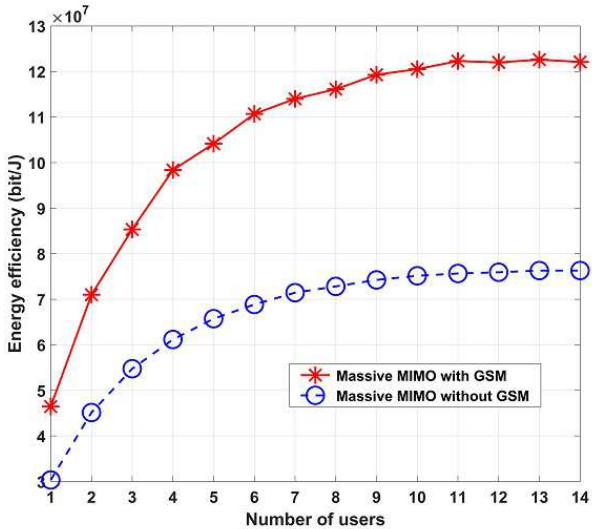}}
\caption{Energy efficiency under different numbers of users.}
\label{fig5}
\end{figure}

\begin{figure}[htbp]
\centerline{\includegraphics[width=0.5\textwidth]{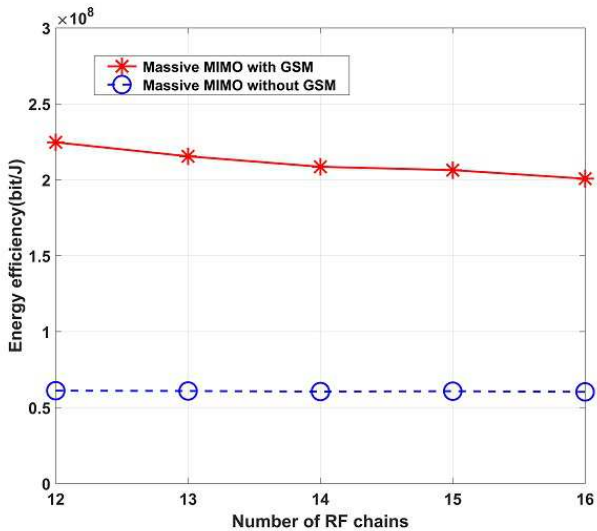}}
\caption{Energy efficiency under different numbers of RF chains.}
\label{fig6}
\end{figure}

Fig. 4 depicts the SE performance under different numbers of users. In Fig. 4, the SE increases with the increase of the number of users. The SE of massive MIMO communication systems with GSM is slightly lower compared to that of conventional massive MIMO communication systems as the number of users is constant,. The EE performance under different numbers of users is illustrated in Fig. 5. In Fig. 5, the EE increases with the increase of the number of users. The EE of massive MIMO communication systems with GSM is maximally improved by approximately 56\% compared with conventional massive MIMO communication systems. The reason is that the total power consumption is greatly reduced while the SE is slightly decreased in massive MIMO communication systems with GSM.

Finally, the EE performance with respect to the number of RF chains (with ${N}_\text{m}=16$, ${N}_\text{k}=8$, ${K}=10$) is depicted in Fig.6. As seen from the figure, when the number of RF chains increases, the EE of massive MIMO communication systems with GSM decreases while the EE of conventional massive MIMO communication systems remains almost constant. Compared with the conventional massive MIMO communication systems, the EE of massive MIMO communication systems with GSM is also improved. Therefore, the higher EE is achieved in massive MIMO communication systems with GSM.

\section{Conclusion}
In this paper, the EE of massive MIMO communication systems with GSM considering the computation power is investigated. Compared with conventional massive MIMO communication systems, simulation results show that the total power consumption and computation power of massive MIMO communication systems with GSM are reduced. As a consequence, the EE of our proposed system is maximally improved by 56\% compared with conventional massive MIMO communication systems. Therefore, the lower computation power and higher EE is achieved by massive MIMO communication systems with GSM.

%\section*{References}
%
%Please number citations consecutively within brackets \cite{b1}. The
%sentence punctuation follows the bracket \cite{b2}. Refer simply to the reference
%number, as in \cite{b3}---do not use ``Ref. \cite{b3}'' or ``reference \cite{b3}'' except at
%the beginning of a sentence: ``Reference \cite{b3} was the first $\ldots$''
%
%Number footnotes separately in superscripts. Place the actual footnote at
%the bottom of the column in which it was cited. Do not put footnotes in the
%abstract or reference list. Use letters for table footnotes.
%
%Unless there are six authors or more give all authors' names; do not use
%``et al.''. Papers that have not been published, even if they have been
%submitted for publication, should be cited as ``unpublished'' \cite{b4}. Papers
%that have been accepted for publication should be cited as ``in press'' \cite{b5}.
%Capitalize only the first word in a paper title, except for proper nouns and
%element symbols.
%
%For papers published in translation journals, please give the English
%citation first, followed by the original foreign-language citation \cite{b6}.

\vspace{12pt}

\end{document}